# A Suggested Alternative to Dark Matter in Galaxies:
# I. Theoretical Considerations

Hanna A. Sabat[1*], Raed Z. Bani-Abdoh[2] and Marwan S. Mousa[2]

*1 Regional Center for Space Science & Technology Education for Western Asia (RCSSTE-WA), Amman, Jordan*

*2 Mu'tah University, Department of Physics, Karak, Jordan*

* Email addresses: sabat@aabu.edu.jo ; hahanna70@gmail.com

**ABTRACT**

Since the dark matter paradigm is not so satisfactory at the galactic scales, we have resorted to a form of Modified Newtonian dynamics (MOND). We have assumed that either (i) the gravitational constant is a function of distance scale, or, (ii) the gravitational-to-inertial mass ratio is a function of distance scale. We have used a linear approximation of each function, from which two new parameters appeared that have to be determined: $G_1$, the first-order coefficient of gravitational coupling, and $C_1$, the first-order coefficient of gravitational-to-inertial mass ratio. We have generated simplified theoretical rotation curves for some hypothetical galaxies by varying our model's parameters. We have concluded that our model gives a qualitatively and quantitatively acceptable behavior of the galactic rotation curves for some values of those parameters: $G_1$ between around $10^{-31}$ to $10^{-30}$ m$^2$ s$^{-2}$ kg$^{-1}$; or, $C_1$ between $10^{-21}$ to $10^{-20}$ m$^{-1}$. Our model also may imply the existence of a critical distance at which the MOND effects become significant rather than a critical acceleration. Furthermore, assuming that the critical centripetal acceleration in our model is equivalent to that in Milgrom's MOND ($a_0$), we found that the $a_0$ is not a constant but a linear function of the galactic baryonic mass ($a_0 \approx G_1^2 M_b / G_0$), and we were able to re-derive Milgrom's version of MOND ($a_c^2/a_0 = G_0 M_b / r^2$).

# 1. INTRODUCTION

Dark matter is the generally accepted paradigm among astrophysicists and cosmologists to explain the higher rate of rotation in galaxies than could be sustained by the amount of the observed normal matter alone, and to explain the higher velocity dispersions of galaxies in galactic clusters than what is observed. For a historical review of the dark matter issue, one may refer to [1, 2] Scientists also consider, in their modern cosmological models, such as the ΛCDM model (which contains dark energy, cold dark matter, and ordinary matter) that dark matter is an essential ingredient that plays a central role in our understanding of the large-scale structure of the Universe as well as of the microwave background radiation. Recent studies based on the Planck mission data, for example, estimate that dark matter adds up to 26.8 percent of the total mass of the Universe, whereas baryonic (ordinary) matter accounts for just 4.9 percent, the rest being dark energy [3] That is, we have more than five times more dark matter than ordinary matter in the Universe.

## 1.1 Galactic Rotation Curves

By the 1970s, it was confirmed that the observed rotation curves for galaxies did not follow the expected Keplerian behavior in the outer regions of galaxies (in the sense that the circular velocities $v$ of stars and/ or gas in the galactic disc is proportional to $r^{-1/2}$, $r$ being the radial distance from the galactic centre); rather, it was found that the speed v almost remains constant after peaking – what is known as flat rotations curves [4] To explain the observations, the majority of researchers were led to accept the fact that galaxies contain unseen dark matter as an important component of galaxies, where the dark matter's mass should increase with radius in order for rotation velocities to remain constant. That is, according to the standard view, most of the dark matter in galaxies is contained in the galactic halos [1].

## 1.2 Some Problems with Dark Matter

Despite the successes of the ΛCDM model in describing the large-scale structure of the Universe, the main problem with dark matter remains that its nature is still unknown. There are many dark matter candidates from particle physics, the most famous of which are the Weakly Interactive Massive Particles, or WIMPs. But the list of dark matter candidates also includes super-WIMPs, light gravitinos, hidden dark matter, sterile neutrinos, and axions [5] But the problem remains that all experiments to detect particles that might be dark matter candidates have given negative results so far. For a review of the current situation of the research on dark matter candidates, (see, for instance: [6, 7]).

There are, of course, other problems with the dark matter paradigm, particularly at the galactic level, such as the core-cusp problem, known as the cuspy halo problem, which refers to a discrepancy between the inferred dark matter density profiles of low-mass galaxies and the density profiles predicted by cosmological N-body simulations (see, for instance: [8]).

**1.3 Some Suggested Alternatives to Dark Matter**

For the above reasons, among other things, many researchers have attempted to find some alternatives to the dark matter paradigm in order to explain observational results, particularly at the galactic scales.

Perhaps the most popular of these attempts is the Modified Newtonian Dynamics (MOND), which was put forward by Milgrom [9, 10, 11] in the early 1980s. Milgrom proposed a modification of Newton's laws to account for the observed properties of galaxies as an alternative to the invisible dark matter halos at the galactic level. His idea is based on the assumption that at accelerations well above a certain critical acceleration, $a_0$: $a/a_0 \gg 1$, Newton's second law applies, whereas at very low accelerations: $a/a_0 \ll 1$, that law becomes: $F = m\, a^2/a_0$, which eventually leads to constant circular velocities at large radial distances from the galactic centre. MOND fully describes the rotation curves of some galaxies given only it is baryonic mass, where it predicts a far strong correlation between the baryonic mass distribution and the dark matter hypotheses [12].

Since Milgrom's original proposal, proponents of MOND have claimed to successfully predict a variety of galactic phenomena that they state are difficult to understand as consequences of dark matter (See, for instance [13, 14]). Bekenstein [15] also suggested a relativistic generalisation of MOND known as the Tensor–Vector–Scalar (TeVeS) Theory. However, MOND and its generalizations do not adequately account for the observed properties of galactic clusters, and no satisfactory cosmological model has been constructed from the MOND hypothesis. On the other hand, many researchers found that the critical acceleration $a_0$ cannot be constant in individual galaxies whose rotation curves were used to obtain its best-fit value (See, for instance: [16, 17]). Furthermore, experiments at extremely low accelerations (below $a_0$) have been conducted, finding no departure from Newton's second law [18].

Moffat [19, 20] suggested a theory of Modified Gravity (MOG), not only to account for galactic rotation curves without invoking dark matter [21, 22] but also as an alternative to dark matter in general and to dark energy on the cosmological scales.

Fahr [23] suggested using a gravitational analogue of the Lorentz force of electromagnetism by introducing a "gravo-inductive" term to the usual "static form" of the force of gravity, which would give rise to flat rotation curves without the need of dark matter. Sivram [24], however, concluded that such gravo-inductive effects are too small to account for flat rotation curves. More recently, Arbab [25] also suggested using a gravitational analogue of the Lorentz force by introducing a "gravito-magnetic" term to gravity in what he called the generalized Newton's law of gravitation, and he compared it to Milgrom's MOND. Altaie & Suleiman [26] suggested the existence of a drag force in the outer regions of spiral galaxies, due to some sort of a dynamically generated viscous medium, which would counterbalance the centripetal force and thus give terminal velocities to stars in those regions.

Other researchers suggested a scale-dependent, or varying, gravitational constant $G$. For instance, Bertolami & Garcia–Bellido [27, 28] argued about the possibility of a scale-dependent gravitational coupling that may have many consequences in astrophysics and cosmology, among which the flatness of galactic rotation curves—though their model required the existence of some dark matter to be compatible with observations. Christodoulou & Kazanas [29], using the baryonic Tully-Fisher and the Faber-Jackson relations, concluded that the gravitational constant $G$ is inversely proportional to acceleration $a$. Vagnozzi and some other researchers used "mimetic gravity" as a tool to obtain MOND-like acceleration laws that might explain flat rotation curves (see, for instance: [30]).

In fact, a scale-dependent gravitational constant (as mentioned above) may provide a theoretical background for the model we are suggesting in this research, as will be shown below.

**1.4 About this Work**

We shall use a new approach to Modified Newtonian Dynamics (MOND). In this part of our work (Part I: Theoretical Considerations), we shall only investigate the theoretical aspects of our model and its implications. We shall start by presenting the physical and mathematical basis of our work (Section 2), in which we detail the mathematical framework of the classical Newtonian dynamics of galactic rotation and our version of MOND. In Section 3, we shall generate some theoretical simplified rotation curves based on our model using computer coding, by varying our model's parameters, and contrasting them to the classical Newtonian approach. Then, we shall discuss all the obtained results in Section 4, and present our conclusions in Section 5.

In the second part of this work (Part II: Observational Considerations), we shall analyze some observational rotation curves for a number of galaxies, and try to evaluate our model's parameters from them.

## 2. THE PHYSICAL-MATHEMATICAL FOUNDATION OF THE MODEL

We shall start here by considering the orbital speed of a star, or a blob of gas, which lies outside the galactic bulge of a galaxy, at a distance $r$ from the galaxy's center (GC). Assuming that this star (or blob of gas) has an inertial mass $m_i$ and a gravitational mass $m_g$, and that it is only influenced by the gravitational pull of the galactic bulge's mass, $M_b$, the orbital speed of the star or blob of gas $v_c$ can be estimated by using Newton's 2nd law as follows:

$$\Sigma F = m_i a_c = m_i \frac{v_c^2}{r}$$

$a_c$ being the centripetal acceleration. Then, using Newton's law of gravitation:

$$F_g = \frac{GM_b m_g}{r^2} = m_i \frac{v_c^2}{r}$$

$$v_c = \left(\frac{GM_b m_g / m_i}{r}\right)^{1/2} \tag{1}$$

Eq. (1) above is the well-known Keplerian speed with its $r^{-1/2}$ dependence, but where we did not cancel the gravitational mass with the inertial mass in the equation, for reasons to be explained below.

In order for the speed, $v_c$, in eq. (1) to divert from the Keplerian behavior, the terms in the numerator should somehow be a function of distance in one of the following ways:

- either the gravitational constant, $G$, is a function of distance: $G = G(r)$, probably as a result of a scale-dependent gravitational coupling, or,
- the gravitational-to-inertial mass ratio, or $m_g/m_i$, is a function of distance, that is: $m_g/m_i = f(r)$, which means that the equivalence principle is somehow violated at some large distance-scales.

However, before using this sort of Modified Newtonian Dynamics (MOND), we shall rederive the basic equations of the Newtonian dynamics in galaxies, since they will be needed in both parts of our work.

### 2.1 Purely Newtonian Behavior

#### 2.1.1 Within the Galactic Bulge

Let us start with a star that lies at a distance $r$ from the galactic center (GC) of a certain galaxy, but within its galactic bulge; the volume of the galactic sphere of radius $r$ that is concentric with the GC is:

$$V_r = \frac{4\pi}{3} r^3$$

The total mass of matter contained within this spherical volume is:

$$M_r = \langle\rho\rangle V_r = \langle\rho\rangle \frac{4\pi}{3} r^3 \qquad (2)$$

Where $\langle\rho\rangle$ is the average mass density of the galactic bulge. That is:

$$\langle\rho\rangle = \frac{M_b}{V_b} = \frac{M_b}{\frac{4\pi}{3} r_b^3} \qquad (3)$$

$M_b$ & $r_b$ being the total mass of the galactic bulge and its radius, respectively.

Assuming that the star lying at a distance $r$ from the galactic center has a mass $m$, it will therefore experience a force of gravity $F_g$ only from the mass contained within the sphere of radius $r$ and mass $M_r$, which is concentric with the galactic center. Using Newton's law of gravitation and eq. (2), we get the gravitational force acting on the star as:

$$F_g = \frac{G M_r m}{r^2} = \frac{G\langle\rho\rangle \frac{4\pi}{3} r^3 m}{r^2}$$

$$F_g = \frac{4\pi}{3} G\langle\rho\rangle \, m \, r \qquad (4)$$

Assuming that stars inside the galactic bulge experience a purely centripetal acceleration due to the influence of gravitation, we may use Newton's 2nd law to calculate the orbital (or circular) speed, $v_c$, of this star:

$$\Sigma F = m \, a_c = m \, \frac{v_c^2}{r} \qquad (5)$$

From the previous two equations (4) & (5), we get:

$$\Sigma F = F_g = m\, a_c = m\, \frac{v_c^2}{r}$$

$$\frac{4\pi}{3} G \langle \rho \rangle\, m\, r = m\, \frac{v_c^2}{r}$$

$$v_c^2 = \frac{4\pi}{3} G \langle \rho \rangle\, r^2$$

$$v_c = \left(\frac{4\pi}{3} G \langle \rho \rangle\right)^{\frac{1}{2}} r$$

$$= k_g\, r \qquad (6)$$

Where:
$$k_g = \left(\frac{4\pi}{3} G \langle \rho \rangle\right)^{1/2} \qquad (7)$$

Noting that $k_g$, the constant of proportionality between the rotational speed and distance r, may vary from one galaxy to another, since it depends on the value of the mean mass density of the galactic bulge, $\langle \rho \rangle$. If $k_g$ is estimated observationally for a given galaxy, the mean mass density $\langle \rho \rangle$ of its galactic bulge may be estimated using eq. (7) as:

$$\langle \rho \rangle = k_g^2 \left(\frac{4\pi}{3} G\right)^{-1} \qquad (8)$$

Furthermore, if the radius of the galactic bulge, $r_b$, is also estimated for the same galaxy, its total mass, $M_b$, may be determined using eq. (3).

The centripetal acceleration in this case, using eq. (6), will be:

$$a_c = \frac{v_c^2}{r} = \frac{(k_g\, r)^2}{r}$$

$$a_c = k_g^2\, r \qquad (9)$$

### 2.1.2 Outside the Galactic Bulge

If a star of mass *m* lies at a distance *r* from the galactic center (GC) of a certain galaxy, but well beyond its galactic bulge, we may assume that this star is mainly influenced by the gravitation of the total mass contained within this galactic bulge, $M_b$, if we neglect the influence of the disc stars. In this case, the gravitational force upon this star will be:

$$F_g \approx \frac{GM_b m}{r^2} \quad (10)$$

Applying Newton's 2nd law to calculate the circular speed $v_c$ of this star and inserting eq. (10):

$$\Sigma F = m\, a_c = m\, \frac{v_c^2}{r}$$

$$\Sigma F = F_g = m\, \frac{v_c^2}{r}$$

$$\frac{GM_b m}{r^2} = m\, \frac{v_c^2}{r}$$

$$v_c = \left(\frac{GM_b}{r}\right)^{1/2} \quad (11)$$

Which is the normal Keplerian speed, with its $r^{-1/2}$ dependence.

### 2.2 Using our Version of Modified Newtonian Dynamics (MOND)

In this work, we shall not use the MOND treatment as proposed by Milgrom [9, 10, 11] which considers that the laws of mechanics deviate from Newtonian mechanics at very low accelerations, rather, we will assume either that the gravitational coupling has a sort of scale-dependence, or that the equivalence between the gravitational and inertial masses has a sort of scale dependence.

### 2.2.1 Using a Scale-Dependent *G*

Motivated by works that considered a scale-dependent gravitational coupling (see, for instance: [27, 28, 29, 30]), we will assume here that the gravitational constant, *G*, has some sort of dependence on the scale considered, that is, it is a function of distance as *G(r)*. But since we do not know the exact dependence of the function *G(r)* on distance, we will use a Taylor series approximation of the function as:

$$G(r) = G_0 + G_1 r + \frac{1}{2} G_2 r^2 + \cdots \tag{12a}$$

Where $G_0$ is the Newtonian constant of gravitation, which we will designate here as the zero-order coefficient of gravitational coupling; $G_1$ is the first-order coefficient of gravitational coupling, $G_2$ is the second-order coefficient of gravitational coupling, and so forth …

However, for our purposes in this research, where we are interested in distance scales within galaxies, we will limit our treatment to the first two terms of the series and neglect higher-order terms, that is:

$$G(r) \approx G_0 + G_1 r \tag{12b}$$

Now, if we consider a star that lies at a distance *r* from the galactic center (GC) of a certain galaxy, but well beyond its galactic bulge, and neglecting the gravitational influence of matter outside the bulge (as a first approximation), the star will be mainly influenced by the mass of the galactic bulge, $M_b$, in which case, using the modified form of gravitational coupling – eq. (12b), the gravitational force upon this star will be:

$$F_g = \frac{G(r) M_b m}{r^2} \approx \frac{(G_0 + G_1 r) M_b m}{r^2} \tag{13}$$

To calculate the circular speed, $v_c$, for this star, we apply Newton's 2nd law and insert eq. (13) into it:

$$\Sigma F = m\, a_c = m\, \frac{v_c^2}{r}$$

$$F_g = m\, \frac{v_c^2}{r}$$

$$\frac{(G_0 + G_1 r)M_b m}{r^2} \approx m \frac{v_c^2}{r}$$

$$v_c \approx \left(\frac{(G_0 + G_1 r)M_b}{r}\right)^{1/2}$$

$$v_c \approx \left(\frac{G_0 M_b}{r} + G_1 M_b\right)^{1/2} \tag{14}$$

The latter equation expresses the modified form of the circular speed $v_c$. But it should be noted here that (using the same system of units) the value of $G_1$ should be many orders of magnitude smaller than the Newtonian constant of gravitation, $G_0$, in order to be consistent with experiments and observations, as will be shown.

At distance scales much smaller than a certain critical distance, $r_c$ (at $r \ll r_c$), the 1st term under the square root in eq. (14) is dominant, so that the circular speed $v_c$ reduces to its Keplerian form with its $r^{-1/2}$ dependence.

At distance scales comparable to that critical distance, $r_c$ (at $r = r_c$), the value of the 2nd term under the square root in eq. (14) becomes comparable to the 1st one, that is:

$$\frac{G_0 M_b}{r_c} \approx G_1 M_b$$

Or:
$$r_c \approx G_0 / G_1 \tag{15}$$

When we consider distance scales that are much larger than the critical distance ($r \gg r_c$), we may neglect the 1st term under the square root in eq. (14), so that the circular speed of a star at such a distance, the asymptotic speed, becomes:

$$v_{c,asym} \approx (G_1 M_b)^{\frac{1}{2}} \approx const. \tag{16}$$

Noting here that we have neglected the possible effects of the coefficient $G_2$ and higher terms in eq. (12a).

At the critical distance, $r_c$, the critical centripetal acceleration can be estimated using eqs. (14) & (15) as follows:

$$a_{c,cr} = \frac{v_{c,cr}^2}{r_c} = \frac{1}{r_c}\left(\frac{G_0 M_b}{r_c} + G_1 M_b\right)$$

$$a_{c,cr} = \frac{G_0 M_b}{r_c^2} + \frac{G_1 M_b}{r_c}$$

$$= \frac{G_0 M_b}{(G_0/G_1)^2} + \frac{G_1 M_b}{G_0/G_1}$$

$$a_{c,cr} = \frac{2 G_1{}^2 M_b}{G_0} \approx (G_1{}^2/G_0) M_b \qquad (17)$$

That is, the critical centripetal acceleration here is proportional to the total (baryonic) mass of the galactic bulge, unlike Milgrom's MOND, where his critical acceleration is assumed to be constant [16].

**2.2.2 Using a Scale-Dependent Gravitational-to-Inertial Mass Ratio**

It is possible to obtain similar results as above if the equivalence principle is violated at large distance scales. Let us start by using eq. (1) to express the circular speed of a star at a distance $r$ from the galactic center (GC), while retaining the ratio $m_g/m_i$:

$$v_c = \left(\frac{GM_b}{r}\frac{m_g}{m_i}\right)^{1/2} \qquad (18)$$

Let us assume here that the gravitational constant does not vary here with distance, as in the classical Newtonian approach, but that the equivalence principle is somehow violated at some large distance scales, i.e., the gravitational and inertial masses are no longer equivalent at such scales, and that the gravitational-to-inertial mass ratio may be expressed as a function of some distance scale: $m_g/m_i = f(r)$. Since that function is unknown to us, we may expand it as a Taylor series approximation as:

$$\frac{m_g}{m_i} = f(r) \approx 1 + C_1 r + C_2 r^2 + \cdots \qquad (19)$$

If we neglect the second-order and higher terms in eq. (19) – as we did with the gravitational coupling function above – and only take the first two terms in the equation, then by inserting it into eq. (18), we get:

$$v_c = \left(\frac{GM_b}{r}[1 + C_1 r]\right)^{1/2}$$

$$v_c = \left(\frac{GM_b}{r} + C_1 G M_b\right)^{1/2} \qquad (20)$$

It should be noted that the value of the coefficient $C_1$ should be very small in order for eq. (20) to reduce to the Keplerian speed at distances much smaller than the critical distance $r_c$. At the critical distance, $r = r_c$, the two terms under square root in eq. (20) become equal, so that:

$$\frac{GM_b}{r_c} \approx C_1 GM_b \Rightarrow r_c \approx 1/C_1 \qquad (21)$$

At very large distances from the galactic center, much larger than the critical distance ($r \gg r_c$), we may neglect the first term under square root in eq. (20), so that the asymptotic circular speed of a star at such a distance becomes:

$$v_{c,asym} \approx (C_1 GM_b)^{\frac{1}{2}} \approx const. \qquad (22)$$

-- neglecting the probable effects of the coefficient $C_2$ and higher terms in eq. (19).

By comparing eq. (16) with eq. (22), i.e., the asymptotic circular speeds of stars (or gas) within galaxies at large distances from the galactic center, we notice that:

$$v_c \approx (G_1 M_b)^{1/2} \approx (C_1 GM_b)^{1/2} = (C_1 G_0 M_b)^{1/2}$$

That is:

$$C_1 = G_1/G_0 = 1/r_c \qquad (23)$$

-- using eq. (21).

**2.3 Relationship with Milgrom's MOND**

To illustrate the relationship between our approach to Modified Newtonian Dynamics (MOND) and Milgrom's MOND – as mainly described in Milgrom [16] [17] [18], let us start with the critical centripetal acceleration we have introduced in eq. (17). By assuming that it is analogous to Milgrom's critical acceleration, or, $a_0 \approx a_{c,cr}$, we may rewrite eq. (17) as:

$$a_0 \approx \frac{G_1^2 M_b}{G_0}$$

From which:

$$G_1 = (a_0 G_0/M_b)^{\frac{1}{2}} \quad (24)$$

Then, applying Newton's 2$^{nd}$ law, and inserting the modified law of gravity – eq. (13), we get

$$\Sigma F = \frac{(G_0 + G_1 r) M_b m}{r^2} = m\, a_c$$

From which, at large enough distances (so that $G_0$ may be neglected), we get:

$$a_c \approx G_1\, M_b/r \quad (25)$$

By inserting $G_1$ from eq. (24) into eq. (25), squaring both sides and rearranging, we get:

$$a_c = \left(\frac{a_0 G_0}{M_b}\right)^{\frac{1}{2}} \frac{M_b}{r}$$

$$a_c^2 = \frac{a_0 G_0}{M_b} \frac{M_b^2}{r^2}$$

$$\frac{a_c^2}{a_0} = \frac{G_0 M_b}{r^2} \quad (26)$$

Where eq. (26) is exactly Milgrom's MOND equation.

Furthermore, starting by eq. (17), we have: $a_0\, G_0 \approx G_1^2 M_b$. By multiplying this equation by the mass $M_b$, we get:

$$a_0\, G_0 M_b \approx G_1^2 M_b^2 = (G_1\, M_b)^2$$

Then, by taking the fourth root of the latter equation, we get:

$$v_{c,asym} \approx (a_0\, G_0 M_b)^{1/4} \approx (G_1\, M_b)^{1/2} \quad (27)$$

Where both parts of eq. (27) give the asymptotic circular speed in the outer regions of the galaxy, $v_{c,asym}$, the middle part using Milgrom's MOND, and the right part using our model – from eq. (16). In fact, the middle part of eq. (27) turns into its right part by simply expressing the critical acceleration as $a_0 = (G_1^2/G_0)\, M_b$, from eq. (17).

## 3. THEORETICAL ROTATION CURVES

As a first approximation, and in order to study the pure effect of our Modified-Newtonian approach, we have assumed that the total galactic mass is centered in the galactic bulge, thus neglecting the effects of mass distribution outside the bulge. The goal here was to see which combination of values of the galactic bulge (baryonic) mass, $M_b$, and the 1st-order coefficient of gravitational coupling, $G_1$, may yield reasonable values of circular speeds for stars or gas outside the galactic bulge.

Based on the above, we shall present below the results of the theoretical (simulated) rotation curves that we have generated for our hypothetical galaxies:

1. Using the purely Newtonian behavior for the inner region of the galaxy (galactic bulge), eq. (6), and for its outer region, eq. (11); and,
2. Using our approach to MOND, eq. (14) or eq. (20).

In Table (1), we list the physical constants and fixed parameters that we have used in our model's calculations. In Table (2), we list the input and output parameters for the case of a galaxy that has a constant number of stars in its galactic bulge (i.e., a constant mass), which was fixed to $10^{10}$ stars, but where we have varied the values of the 1st-order coefficient of gravitational coupling, $G_1$ – as shown in the same Table. In Table (3), we list the input and output parameters for the opposite case: a galaxy that has a constant value of the 1st-order coefficient of gravitational coupling: $G_1 = 1 \times 10^{-30}$ m$^2$ s$^{-2}$ kg$^{-1}$, but where the number of stars in the galactic bulge (or, its total mass) was varied as shown in the same Table. Figures (1) to (7) represent the theoretical rotation curves for each case, using both classical Newtonian dynamics and our model.

# 4. DISCUSSIONS

**4.1 Table (2)**

Referring to Table (2), for the case of a galaxy with a constant number of stars in the galactic bulge (i.e., constant mass): NG=$10^{10}$, but using different 1st-order coefficients of gravitational coupling, $G_1$, it is noticed that:

- In general, the critical radius $r_c$ increases with decreasing the value of the coefficient $G_1$, because there is an inverse relationship between them.

- In Case I, when $G_1$ has the largest value ($G_1 = 10^{-29}$ m² s⁻² kg⁻¹), the critical radius of the galactic bulge $R_c$ will be 0.21 kpc. But according to observational results (as will be shown in Part II of this study), this value is too small for ordinary galaxies.

- In Case IV, when $G_1$ has the smallest value ($G_1 = 10^{-32}$ m² s⁻² kg⁻¹), the value of $r_c$ was around 215 kpc, which is too high for real galaxies.

- In general, the values of the mean density of the galactic bulge, $<\rho>$, decrease with decreasing the value of $G_1$, which is expected since the number of stars is constant (the mass is constant), whereas the radius of the galactic bulge increases (i.e., the volume of the galactic bulge increases).

- The values of the proportionality constant $k_g$ decrease with decreasing the value of the coefficient $G_1$, which is also expected, since the mean density $<\rho>$ is decreasing in this case, and the constant $k_g$ is proportional to the mean density as: $k_g \propto (<\rho>)^{1/2}$.

- For every one order-of-magnitude decrease in the value of $G_1$, there are three orders-of-magnitude decrease in the value of $<\rho>$, which is also expected, since the density (at constant number of stars or bulge mass) is inversely proportional with cubic $r_c$: $<\rho> \propto r_c^{-3}$.

- The values of the first-order gravitational-to-inertial mass ratio $C_1$ decrease with decreasing the value of $G_1$, which is expected, since $C_1 = G_1/G_0$, and $G_0$ is constant.

- In general, the values of the critical acceleration, $a_{c,cr}$, decrease with decreasing the value of $G_1$, which is expected since the number of stars is constant (the mass is constant), whereas the radius of the galactic bulge increases:

$$a_{c,cr} = \frac{G_0 M_b}{r_c^2} = \frac{2 G_1^2 M_b}{G_0}.$$

**4.2 Table (3)**

Referring to Table (3), for the case of a galaxy with a constant value of the 1st-order coefficient of gravitational coupling, $G_1 = 1\times10^{-30}$ m² s⁻² kg⁻¹, using different numbers of stars in the galactic bulge, NG (i.e., different masses), it is noticed that:

- The values of the critical radius $r_c$ are constant throughout the three cases, since the 1st-order coefficient of gravitational coupling $G_1$ is constant, with a value of around 2.2 kpc, which is of the same order of magnitude as the observational values, as will be shown in Part II of this study.

- Values of the mean density $<\rho>$ increase with increasing the number of stars in the bulge NG (i.e. increasing the galactic bulge mass), which is also expected, since the galactic bulge radius $r_c$ is constant (i.e., the volume is constant), whereas the mass (number of stars in the bulge) is increasing.

- The values of the proportionality constant $k_g$ increase with increasing the number of stars in the bulge NG (i.e. increasing the galactic bulge mass), which is also expected, since the mean density RHOBAR is increasing in this case, and the constant $k_g$ is proportional to the mean density as: $k_g \propto (<\rho>)^{1/2}$.

- The values of the first-order gravitational-to-inertial mass ratio $C_1$ are constant throughout the three cases, since the 1st-order coefficient of gravitational coupling $G_1$ is constant.

- The values of the critical acceleration $a_{c,cr}$ increase with increasing the value of the number of stars (i.e., bulge mass), which is expected since the radius of the galactic bulge is constant, whereas the number of stars is increasing (the mass is increasing), where: $a_{c,cr} = \frac{G_0 M_b}{r_c^2} = \frac{2 G_1^2 M_b}{G_0}$.

### 4.3 Figures

It should be noted at the outset that all the generated rotation curves have a sharp edge at their maximum values of circular speed, which is unlike the observational rotation curves. This is due to the fact that in our model, we assumed that all the stars are concentrated inside the galactic bulge, which has a spherical shape of a specific (critical) radius and with a constant (average) mass density everywhere, which is not really realistic, of course. But this is done as a first approximation, since our goal here was not to fit observational data with our theoretical model, but rather, to see the general qualitative and quantitative behavior of the generated theoretical rotation curves based on our model's assumptions alone.

Referring to the theoretical rotation curve for Case I in Figure (1), it is noticed that the values of the circular speed are very high compared to observational galactic rotation curves – as will be shown in Part II of this study. For example, the maximum speed is around 400 km/s in the purely Newtonian curve, and around 600 km/s in the modified Newtonian curve. Also, the value of the critical radius is too small (around 0.2 kpc).

Referring to the theoretical rotation curves for Case II and Case III in Figures (2) and (3), respectively, it is noticed that when the number of stars in the galactic bulge is around NG=$10^{10}$ stars, and the values of the 1st-order coefficient of gravitational coupling $G_1$ is between $10^{-30}$ and $10^{-31}$ m² s⁻² kg⁻¹, the values of rotational velocities are compatible with observational values (between around 200 km/s to around 60 km/s); and so are the values of the critical radius (galactic bulge radius), which fall between around 2 kpc to around 20 kpc.

Referring to the theoretical rotation curve for Case IV in Figure (4), where NG=$10^{10}$ stars, and the value of $G_1$ is $10^{-32}$ m² s⁻² kg⁻¹, it is noticed that the values of the circular speed are very low (no more than 20 km/s). Also, the value of the critical radius is too large (around 200 kpc), which has never been observed in any published literature.

Referring to the theoretical rotation curves for Case V and Case VI in Figures (5) and (6), respectively, it is noticed that when we have a constant value of the 1st-order coefficient of gravitational coupling $G_1$=$10^{-30}$ m² s⁻² kg⁻¹, and a number of stars in the galactic bulge between NG = $10^9$ to $10^{10}$, values of the theoretical rotational velocities (60 km/s – 200 km/s) are compatible with observational values, but with a constant value of the critical radius (bulge radius) of around 2 kpc.

Referring to the theoretical rotation curve for Case VII in Figure (7), where the value of $G_1=10^{-30}$ m$^2$ s$^{-2}$ kg$^{-1}$ and the number of stars was increased to $10^{11}$ (with a galactic bulge radius of 2 kpc), it is noticed that the values of the circular speed are too high (up to around 600 km/s), which may not be compatible with observations.

Considering all the theoretical rotation curves, it is noticed that for the inner region of the galaxy (i.e., the galactic bulge), the modified Newtonian curve is a bit higher than its corresponding purely Newtonian curve, and a little bit curved rather than linear.

## 5. CONCLUSIONS

Based on the above results and discussions, we may conclude the following:

- Using our version of Modified Newtonian Dynamics (MOND), which consists of either: using a linearized function of the scale-dependent gravitational coupling: $G(r)$; or, using a linearized function of the scale-dependent gravitational-to-inertial mass ratio $f(r)$; the model gives a qualitatively and quantitatively acceptable behavior of the galactic rotation curves for certain values of the model's parameters.

- At very small distance scales, either of the following occurs: the effect of the first order coefficient of the scale-dependent gravitational coupling $G_1$ becomes negligible, and we turn back to classical Newtonian dynamics, or, the effect of the first order coefficient of the scale-dependent gravitational-to-inertial mass ratio $C_1$ becomes negligible, and we turn back to classical Newtonian dynamics.

- The values of the first-order coefficients that give quantitatively acceptable description of galactic rotation curves are as follows: For the first-order coefficient of gravitational coupling $G_1$, it falls between around $10^{-31}$ to $10^{-30}$ m$^2$ s$^{-2}$ kg$^{-1}$. As for the first-order coefficient of gravitational-to-inertial mass ratio $C_1$, it falls between around $10^{-21}$ to $10^{-20}$ m$^{-1}$. Both values should be further refined by comparison to observations – which is left to Part II of this study.

- Relationship with Milgrom's MOND: Assuming that Milgrom's critical acceleration [16] is the same as the critical acceleration we estimated from our model ($a_0 \approx a_{c,cr}$), we conclude the following:
    - The critical acceleration and the 1st order coefficient of gravitational coupling are related thus:
    
    $a_0 \approx \frac{G_1^2 M_b}{G_0}$ – eq. (17').
    
    - Milgrom's equation of Modified Newton's Dynamics (MOND) can be derived: $\frac{a_c^2}{a_0} = \frac{G_0 M_b}{r^2}$ – eq. (26).
    
    - The asymptotic circular speed in the outer regions of the galaxy can be expressed in terms of the critical acceleration, or, in terms of the 1st order coefficient of gravitational coupling as:
    
    $v_{c,asym} \approx (a_0 G_0 M_b)^{1/4} \approx (G_1 M_b)^{1/2}$ – eq. (27).
    
    - Unlike Milgrom's version of Modified Newtonian Dynamics (MOND) that requires a fixed value of critical acceleration to reach before the MOND effects play a significant role (of the order of $10^{-10}$ m$^2$ s$^{-1}$), our version of MOND implies that there is a critical distance at which the MOND effects become significant and that the critical acceleration $a_0$ is proportional to the baryonic mass of the galaxy – according to eq. (17').
    
    - However, if Milgrom's critical acceleration $a_0$ is indeed a constant, then the 1st order coefficient of gravitational coupling $G_1$ depends on the baryonic mass of the galaxy, according to the following formula: $G_1 = (a_0 G_0 / M_b)^{\frac{1}{2}}$ – eq. (24).

- Our version of MOND may be further extended to deal with the scale of galactic clusters and superclusters, up to the cosmological scales. But, in order to deal with such large scales, we need to estimate the higher-order coefficients of the scale-dependent gravitational coupling (or the scale-dependent gravitational-to-inertial mass ratio), which is left for future studies.

| | Table (1): | | |
|---|---|---|---|
| **Fixed Constants and Parameters Used in the Model's Calculations** | | | |
| **Constant/ Parameter** | **Symbol** | **Value** | **Unit** |
| Newtonian Gravitational Constant | G0 | 6.674E-11 | N m$^2$ kg$^{-2}$ = m$^3$ s$^{-2}$ kg$^{-1}$ |
| Solar Mass | MS | 2.000E+30 | kg |
| Light-Year | LY | 9.500E+15 | m |
| Kilo-Parsec | KPC | 3.26E+03 | light-year |

**Table (2):**
**Input and Output Parameters for the Case of a Galaxy with a Constant number of Stars in the Galactic Bulge (Constant Mass): NG=1.000E+10, Using Different 1st-order Coefficients of Gravitational Coupling, G1**

| | **Case I** | **Case II** | **Case III** | **Case IV** |
|---|---|---|---|---|
| **Input Parameters** | G1$^i$ = 1.000E-29 | G1 = 1.000E-30 | G1 = 1.000E-31 | G1 = 1.000E-32 |
| | NG$^{ii}$ = 1.000E+10 | NG = 1.000E+10 | NG = 1.000E+10 | NG = 1.000E+10 |
| | RG$^{iii}$ = 9.500E+20 | RG = 9.500E+20 | RG = 4.750E+21 | RG = 2.850E+22 |
| **Output Parameters** | RC1$^{iv}$ = 6.674E+18 RC2$^{iv}$ = 2.153E-01 | RC1 = 6.674E+19 RC2 = 2.153 | RC1 = 6.674E+20 RC2 = 2.153E+01 | RC1 = 6.674E+21 RC2 = 2.153E+02 |
| | RHOBAR1$^v$ = 1.606E-17 RHOBAR2 = 2.39E+11 | RHOBAR1 = 1.606E-20 RHOBAR2 = 2.39E+08 | RHOBAR1 = 1.606E-23 RHOBAR2 = 2.39E+05 | RHOBAR1 = 1.606E-26 RHOBAR2 = 2.39E+02 |
| | KG1$^{vi}$ = 6.701E-14 KG2$^{vi}$ = 2.08E+03 | KG1 = 2.119E-15 KG2 = 6.57E+01 | KG1 = 6.701E-17 KG2 = 2.08 | KG1 = 2.119E-18 KG2 = 6.57E-02 |
| | C1$^{vii}$ = 1.498E-19 | C1 = 1.498E-20 | C1 = 1.498E-21 | C1 = 1.498E-22 |
| | ACR$^{viii}$ = 2,997E-08 | ACR = 2,997E-10 | ACR = 2,997E-12 | ACR = 2,997E-14 |
| **Output Rotation Curves** | **Figure (1)** | **Figure (2)** | **Figure (3)** | **Figure (4)** |

(i) G1: 1st-order coefficient of gravitational coupling, in m$^2$ s$^{-2}$ kg$^{-1}$.
(ii) NG: number of stars in the galactic bulge.
(iii) RG: total galactic radius, in m.
(iv) RC: critical radius, (1) in m; (2) in kpc.
(v) RHOBAR: mean density of the galactic bulge, (1) in kg m$^{-3}$; (2) in M$_\odot$ kpc$^{-3}$.
(vi) KG: constant of proportionality between speed and distance, (1) in m s$^{-1}$ m$^{-1}$; (2) in km s$^{-1}$ kpc$^{-1}$.
(vii) C1: 1st-order coefficient of gravitational-to-inertial mass ratio, in m$^{-1}$.
(viii) ACR: critical acceleration, in m s$^{-2}$.

Table (3):
Input and Output Parameters for the Case of a Galaxy
with a Constant Value of the 1st-order Coefficient of Gravitational Coupling, G1=1.000E-30, Using Different
Numbers of Stars in the Galactic Bulge, NG (Different Masses)

|  | Case V | Case VI | Case VII |
|---|---|---|---|
| **Input Parameters** | NG[i] = 1.000E+09<br>G1[ii] = 1.000E-30<br>RG[iii] = 9.500E+20 | NG = 1.000E+10<br>G1 = 1.000E-30<br>RG = 9.500E+20 | NG = 1.000E+11<br>G1 = 1.000E-30<br>RG = 9.500E+20 |
| **Output Parameters** | RC1[iv] = 6.674E+19<br>RC2[iv] = 2.153<br>RHOBAR1[v] = 1.606E-21<br>RHOBAR2[v] = 2.39E+07<br>KG1[vi] = 6.701E-16<br>KG2[vi] = 2.08E+01<br>C1[vii] = 1.498E-20<br>ACR[viii] = 2,997E-11 | RC1 = 6.674E+19<br>RC2 = 2.153<br>RHOBAR1 = 1.606E-20<br>RHOBAR2 = 2.39E+08<br>KG1 = 2.119E-15<br>KG2 = 6.57E+01<br>C1 = 1.498E-20<br>ACR = 2,997E-10 | RC1 = 6.674E+19<br>RC2 = 2.153<br>RHOBAR1 = 1.606E-19<br>RHOBAR2 = 2.39E+09<br>KG1 = 6.701E-15<br>KG2 = 2.08E+02<br>C1 = 1.498E-20<br>ACR = 2,997E-09 |
| **Output Rotation Curves** | **Figure (5)** | **Figure (6)** | **Figure (7)** |

(i)      G1: 1st-order coefficient of gravitational coupling, in $m^2 \, s^{-2} \, kg^{-1}$.
(ii)     NG: number of stars in the galactic bulge.
(iii)    RG: total galactic radius, in m.
(iv)    RC: critical radius, (1) in m; (2) in kpc.
(v)     RHOBAR: mean density of the galactic bulge, (1) in $kg \, m^{-3}$; (2) in $M_\odot \, kpc^{-3}$.
(vi)    KG: constant of proportionality between speed and distance, (1) in $m \, s^{-1} \, m^{-1}$; (2) in $km \, s^{-1} \, kpc^{-1}$.
(vii)    C1: 1st-order coefficient of gravitational-to-inertial mass ratio, in $m^{-1}$.
(viii)   ACR: critical acceleration, in $m \, s^{-2}$.

Figure (1): Theoretical Rotation Curve, Case I.

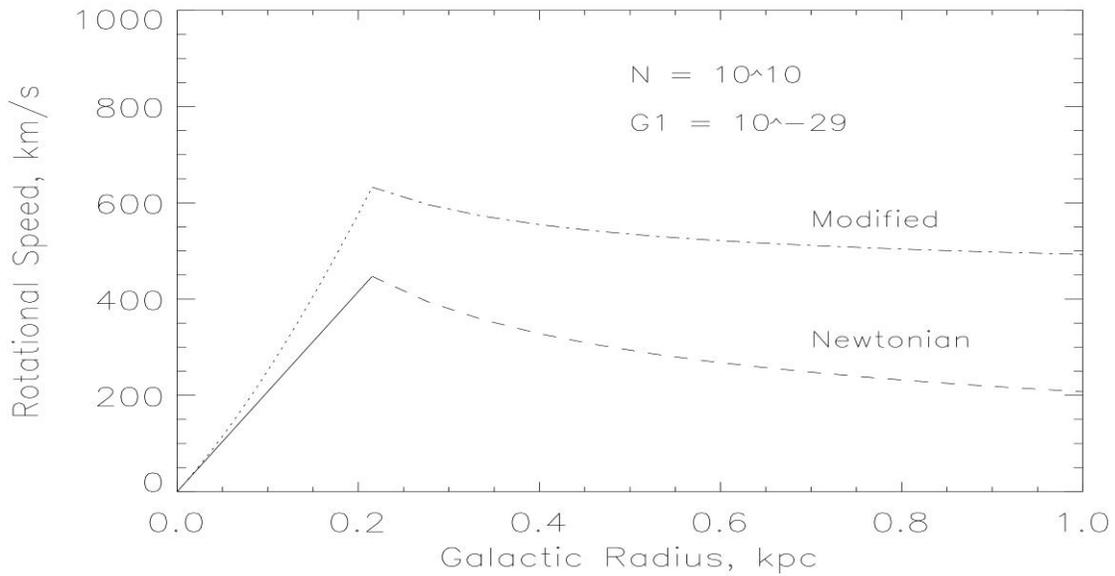

Figure (2): Theoretical Rotation Curve, Case II.

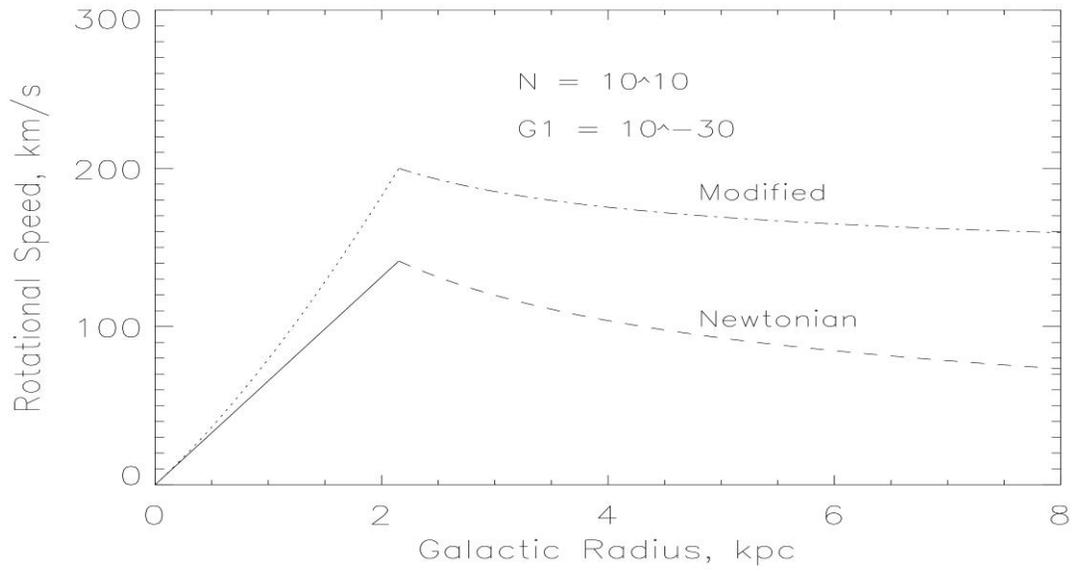

Figure (3): Theoretical Rotation Curve, Case III.

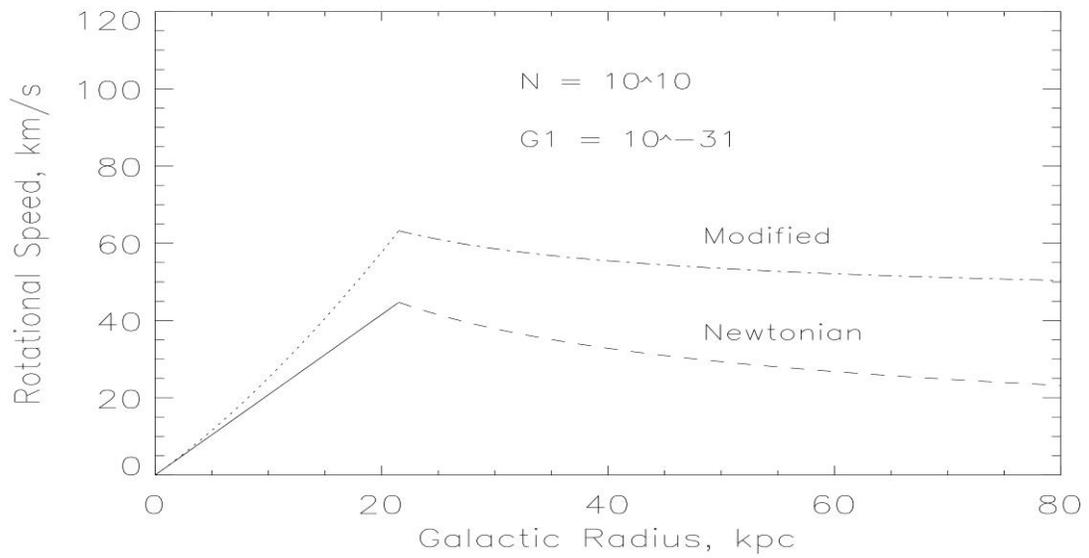

Figure (4): Theoretical Rotation Curve, Case IV.

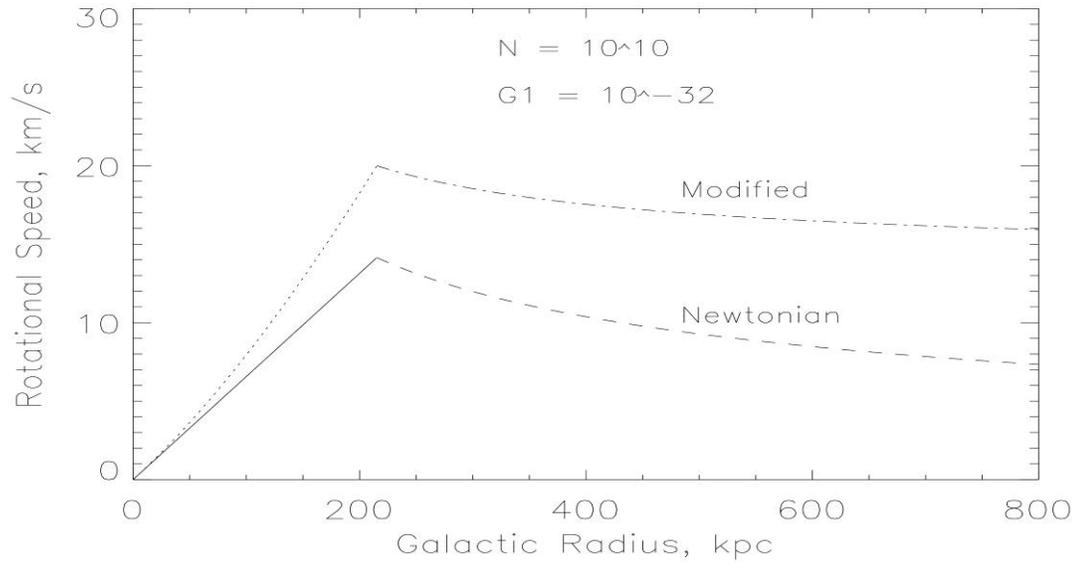

Figure (5): Theoretical Rotation Curve, Case V.

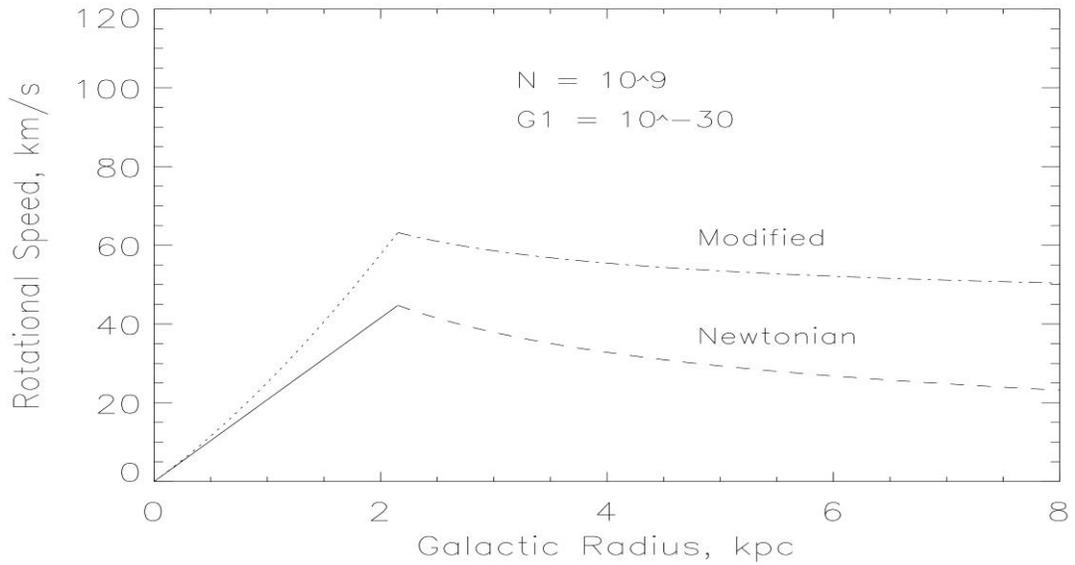

Figure (6): Theoretical Rotation Curve, Case VI.

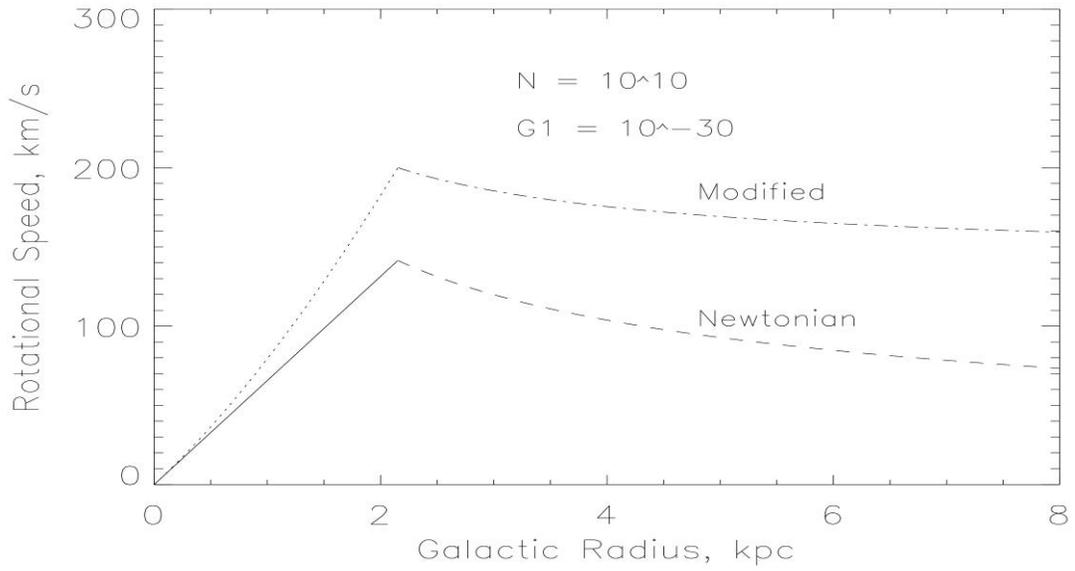

Figure (7): Theoretical Rotation Curve, Case VII.

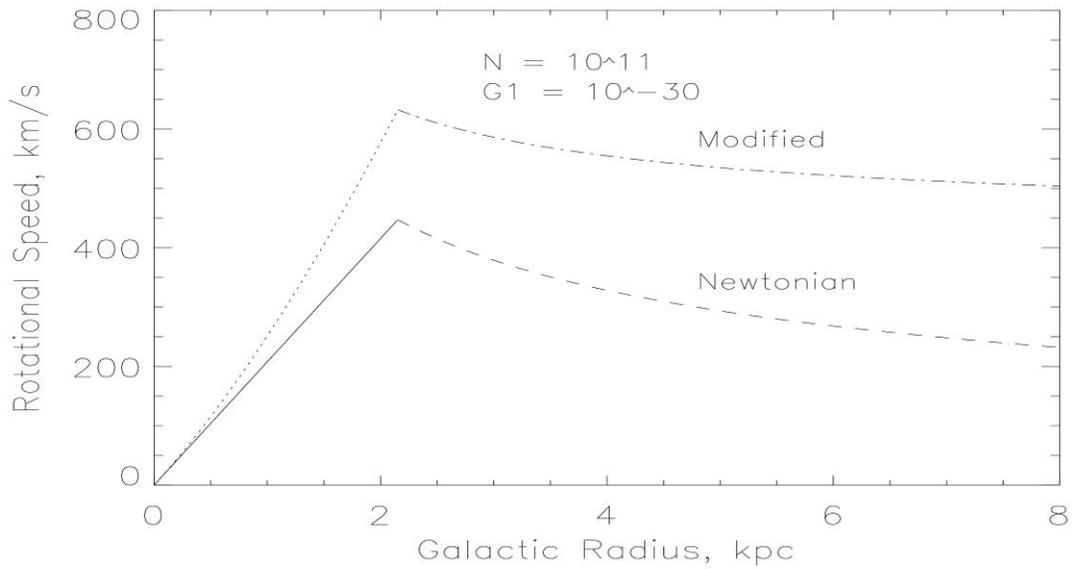